# Timing analysis of a sample of five cataclysmic variable candidates observed by the *XMM*-Newton satellite

A.A. Nucita[1,2,3]* S.M. Lezzi[6,7] F. De Paolis[1,2,3] F. Strafella[1,2,3] D. Licchelli[4,5]
A. Franco[1,2,3] M. Maiorano[1,2,3]

[1] Department of Mathematics and Physics "E. De Giorgi" , University of Salento, Via per Arnesano, CP-I93, I-73100, Lecce, Italy
[2] INFN, Sezione di Lecce, Via per Arnesano, CP-193, I-73100, Lecce, Italy
[3] INAF, Sezione di Lecce, Via per Arnesano, CP-193, I-73100, Lecce, Italy
[4] R.P. Feynman Observatory, I-73034, Gagliano del Capo, Lecce, Italy
[5] CBA, Center for Backyard Astrophysics - I-73034, Gagliano del Capo, Lecce, Italy
[6] University of Naples Federico II, C.U. Monte Sant'Angelo, Via Cinthia, I-80126, Naples, Italy
[7] INAF, Astronomical Observatory of Capodimonte, Salita Moiariello 16, I-80131, Naples, Italy

**ABSTRACT**
Intermediate polars are a class of cataclysmic variables in which a white dwarf accretes material from a companion star. The intermediate polar nature confirmation usually derives from the detection of two periods in both *X*-ray and optical photometry. In this respect, the high energy signal is often characterized by modulations on the white dwarf spin and the orbital period. However, noting that the periodograms may be characterized by strong features also at the synodic period and/or other sidebands, the timing analysis of the *X*-ray signal may offer the unique possibility to firmly discover an intermediate polar candidate. Here, we concentrate on a sample of five cataclysmic variable binary candidates: i.e. SAXJ1748.2-2808, 1RXS J211336.1+542226, CXOGC J174622.7-285218, CXOGC J174517.4-290650, and V381 Vel, listed in the *IPHome* catalogue. Our main aim is to confirm if they belong to the intermediate polar class or not. The results of our analysis show that we can safely assess the intermediate polar nature of all the considered sources, apart for the case of V381 Vel which instead behaves like a cataclysmic variable of the polar subclass. Moreover, the source SAXJ1748.2-2808, previously classified as a HMXB, appears to be, most likely, an intermediate polar variable.

**Key words:** (stars:) novae, cataclysmic variables; X-rays: binaries; X-rays: individual: ..., ..., ...., ...., ....; (stars:) white dwarfs

## 1 INTRODUCTION

A cataclysmic variable (CV) is a binary system constituted by a white dwarf (WD) which accretes material from a companion star and, depending on the details of the accretion mechanism, these objects may be classified in dwarf novae, intermediate polars and polars (see, e.g. Warner 1995; Kuulkers et al. 2006). In particular, dwarf novae are characterized by a WD feeded by a keplerian disk (see, e.g., van Teeseling et al. 1996, Hoard et al. 2010, Nucita et al. 2009, 2011, 2014, Mukai et al. 2017 and reference therein) while the accretion in intermediate polars (see, e.g., de Martino et al. 2004, 2005,Evans & Hellierr 2007) and polars (see, e.g. Ramsay et al. 2004, Szkody et al. 2004) is driven by a moderately large (0.1 − 10 MG) and strong (≳ 10 MG) magnetic field, respectively.

In this CV zoo, intermediate polars are discovered by using a multi-wavelength approach (see, e.g., Patterson et al. 1994 and,

more recently, Wörpel et al. 2020), i.e. observing the same target in the optical and *X*-ray bands. Therefore, possible IP candidates are discovered by using optical surveys such as the Sloan Digital Sky Survey (Gunn et al. 2006) characterizing the prominent emission lines and the shape of the high energy spectrum. The confirmation of the IP nature for these sources comes from the detection of periodic features in the optical as well as *X*-ray light curves. The moderately large magnetic field characterizing IPs makes these binary systems not synchronized and this implies the possible appearance of modulations (Parker et al. 2005) on the WD spin $P_{spin}$, the orbital period[1] $P_{orb}$, the synodic ($P_{syn}^{-1} = P_{spin}^{-1} - P_{orb}^{-1}$) and beat periodicities plus a series of sidebands due to the superpositions of the signal frequencies[2]. The appearance of such features is ultimately due to the existence of different *X*-ray photons reprocessing sites (see, e.g., Warner 1995 for an illustrative case).

---

* E-mail: nucita@le.infn.it

[1] For the intermediate polars class it was shown that the ratio between the WD spin and the orbital period is (see, e.g., Kuulkers et al. 2006) $P_{spin}/P_{orb} = 0.001-1$.

Therefore, recognizing the existence of multiple periodic components (orbital, spin periods, and associated multiple sidebands) in the periodogram of *X*-ray data has been demonstrated to be a valid tool to uncover the IP nature (see, e.g., the discussion hour and case studies in Nucita et al. 2020, 2021 and Wörpel et al. 2020 for recent IP confirmations by this method).

In this respect, long-duration *X*-ray observation with high-sensitivity instruments, as that offered by the cameras on-board the *XMM*-Newton satellite (Jansen et al. 2001), towards intrinsically faint objects are necessary in order to have light curves for which a solid timing analysis can be performed.

Searching and classifying IPs is not only important by its own but also because such objects are expected to be fairly common within our Galaxy and possibly contribute to the *X*-ray ridge background (Worrall et al. 1982). In particular, as Chandra *X*-ray data revealed (see, Revnivtsev et al. 2009), ≃ 80% of the detected *X*-ray flux can be resolved in many faint point sources (including IPs) which seems to dominate the signal above ≃ 10 keV (Warwick et al. 2014). Despite the observational efforts, many IP candidates need a firm confirmation (see e.g. the most updated IP catalogue -*IPhome*- available at https://asd.gsfc.nasa.gov/Koji.Mukai/iphome/iphome.html) and, therefore, dedicated follow-up observation is necessary.

Here we continue the IP confirmation program (see, e.g., Nucita et al. 2020, 2021) of selected targets present in the *IPhome* database and concentrate on the IP candidates SAXJ1748.2-2808, 1RXS J211336.1+542226, CXOGC J174622.7-285218, CXOGC J174517.4-290650, and V381 Vel.

For the each source with coordinates given in the *IPhome* database we searched for a *X*-ray counterpart in the 4XMM-DR11 catalogue (Webb et al. 2020) within 10″. Then, from the resulting observation ID list, we extracted the *XMM*-Newton data with pointing corresponding with the least distance to the source nominal position. This ensures that, for the selected source, it falls in the central region of the telescope field of view. We checked that the duration of the observation was long enough to allow a good periodic analysis. This procedure returned a single observation for 1RXS J211336.1+542226 and V381 Vel, while for the other sources we got several further observations that were not used in the present paper as the main aim was that to confirm the IP nature of the sources. The use of the other observations, allowing the construction of the historical light curve, the study of any period change with time and the spectral analysis, is demanded to a following paper.

We focused on the timing analysis via the Lomb-Scargle technique and address the significance of each feature appearing in the periodogram by following the method described in Lomb (1976) and Scargle (1982), i.e. by comparing the height of each peak with a given threshold corresponding to a false alarm probability in presence of white noise. In constructing the periodogram for each light curve, we tested a number of independent frequencies following the receipt in Horne & Baliunas (1986) and by over-sampling each frequency by a factor three. Furthermore, we confirm (or disprove) the IP nature, by using the method described in Nucita et al. (2021). In particular, we searched for possible signatures of the WD spin, orbital period as well as the associated sidebands in the $0.2 − 10$ keV

*X*-ray light curves and checked whether their relative distance is as expected in the IP accretion scenario. In case the latter condition is satisfied, we consider significant a peak also if it is formally below a given level of significance.

The IP nature is confirmed for all the sources apart from V381 Vel which, according to this analysis, seems to resemble the typical behaviour of CV of the polar subclass.

## 2 TARGET SELECTION AND DATA REDUCTION

As described above, the sources analyzed in this work (see Table 1) were extracted from the *IPhome* database and, for each interesting source, we retrieved the available *XMM*-Newton observations from the *XSA* archive[3].

### 2.1 SAXJ1748.2-2808

The source SAXJ1748.2-2808 (located at J2000 coordinates RA = $17^h48^m16.91^s$ and DEC = $−28°07′59.5″$) was observed by the *XMM*-Newton satellite (ObsID 020524010) for a nominal observing time of ≃ 51 ks. The target was observed by all the instruments on board the satellite and, in particular, with the MOS 1, MOS 2 and pn cameras operating in full frame mode and with the thin filter on. Flares affected ∼ 35% of the entire duration of the observation, which was then cleaned up, as can be seen from the light curves in Figure 2.

The source was discovered by the BeppoSAX towards the Galactic center (GC) region (Sidoli et al. 2001), and first recognized as possibly associated with a protostar or to a giant molecular cloud core. The nature of SAX J1748.2-2808, its very intense Fe K line emission, together with its highly absorbed spectrum (Sidoli et al. 2001), make it a unique object in the GC region. Therefore, the original *XMM*-Newton observation was performed in order to discriminate between different possible origins for the X-ray emission from this interesting source. In fact, according to the spectral analysis performed on the *XMM*-Newton data, Sidoli et al. (2006) concluded that the source could be a low-luminosity high-mass X-ray binary located in the Galactic center region. The discovery of a ≃ 593 s pulsation (Nobukawa et al. 2006) found in Suzaku and *XMM*-Newton data and a flat spectrum characterized by strong emission lines at 6.4–7.0 keV (as due to 3 K-shell lines from neutral and highly ionized irons) make SAX J1748.2-2808 a possible intermediate polar candidate although the high mass X-ray binary scenario cannot be firmly excluded.

### 2.2 1RXS J211336.1+542226

The source 1RXS J211336.1+542226 (also known as SWIFT J2113.5+5422) is located at J2000 coordinates RA = $21^h13^m35.38^s$ and DEC = $+54°22′32.8″$ and was observed by the *XMM*-Newton satellite (ObsID 0761120801) with an exposure time of ≃ 39.3 ks with MOS 1, MOS 2 and pn cameras in full frame mode and thin filter. The observation was affected by strong flares for ∼ 5% of the exposure time in the MOS1 and MOS2 cameras and for almost 53% in the pn camera. Therefore, it was necessary to exclude these data from the analysis, and we focused on the remaining ones, as can be seen from the averaged light curves in Figure 3.

---

[2] Conversely to IP, polars show periodograms characterized by a single prominent periodic feature.

[3] The *XSA* archive is available at the link: https://www.cosmos.esa.int/web/xmm-newton/xsa.

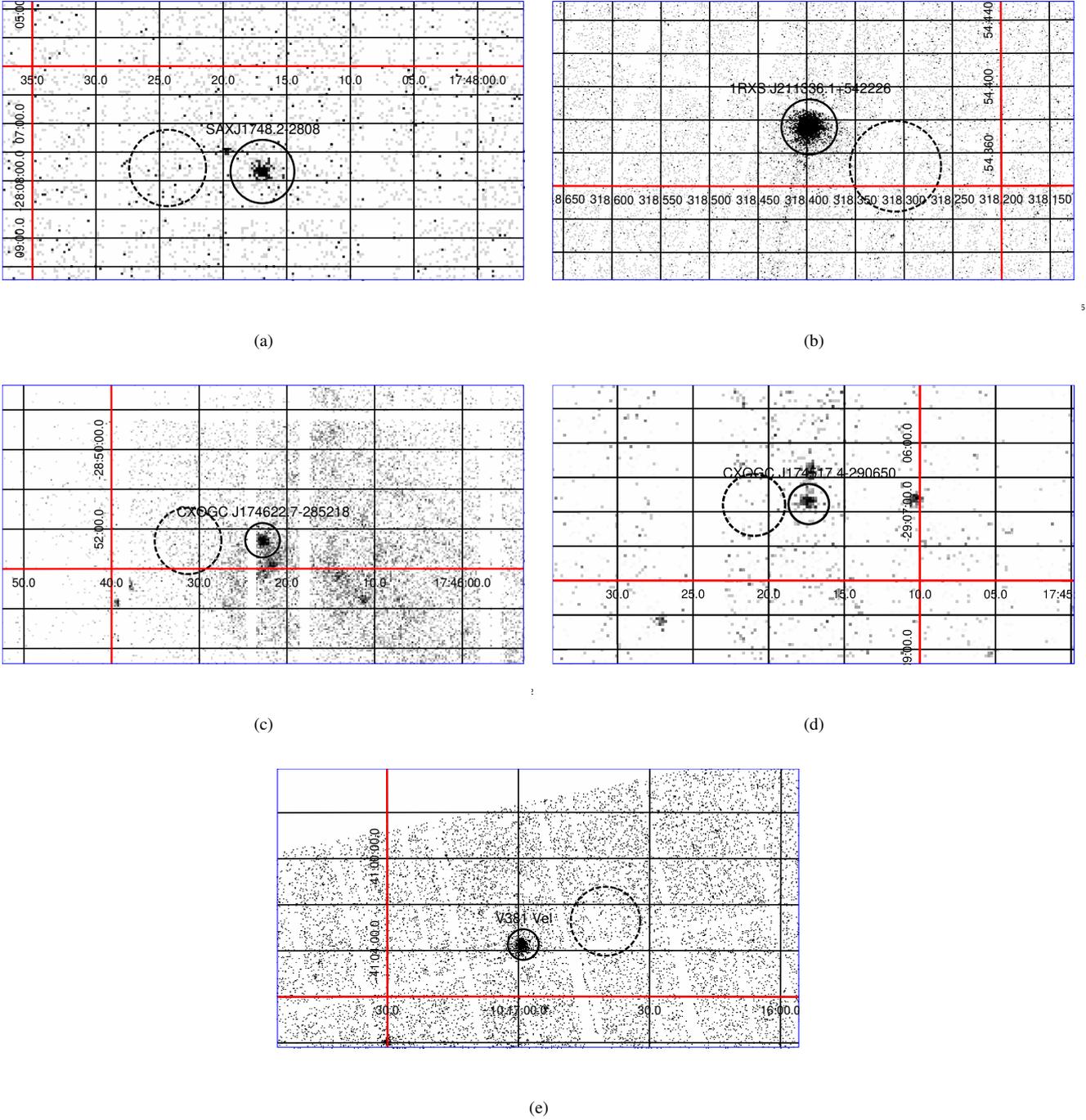

**Figure 1.** Only for illustrative purposes, we give the pn images in the $0.2 − 10$ keV band extracted as described in the text. In particular, in each panel we give a zoom around the nominal target position. The solid circle represents the source plus background extraction region, while the background was estimated using the counts in the dashed circular area.

The target was already catalogued in an Hard X-ray surveys performed by the INTEGRAL satellite and classified by Masetti et al. (2010) as a dwarf nova CV because of the characteristics of its optical spectra although a tentative IP classification can not be excluded based on the observed strength of the *He* II emission line.

The *IPhome* database does not contain any information about the possible WD spin and orbital period although the source was already analyzed by Bernardini et al. (2017). In particular, these authors found evidences of the WD spin and beat period of $21.09 \pm$

0.24 minutes and $22.98 \pm 0.04$ minutes, respectively. Therefore, they expect a peak (corresponding to the orbital period) at $4.46 \pm 0.10$ hours, but found a feature at $4.02 \pm 0.10$ hours instead. Since these two numbers appear to be consistent within $3\sigma$, they confirm the IP nature of 1RXS J211336.1+542226. Note that the source 1RXS J211336.1+542226 was also observed in the optical band by Halpern et al. (2018) who found a period of $4.17 \pm 0.10$ hours still consistent with that determined by Bernardini et al. (2017). However, since no time-resolved optical spectroscopy of Swift J2113.5+5422 is currently available, this period cannot be firmly confirmed.

| Target | ObsID | Time (ks) | Rate (counts s⁻¹) | RA (J2000) | DEC. (J2000) | $P_{spin}$ (s) | $P_{orb}$ (h) |
|---|---|---|---|---|---|---|---|
| SAXJ1748.2-2808 | 0205240101 | 51.0 | 0.024 | $17^h48^m16.91^s$ | $-28°07'59.5''$ | 593 | – |
| 1RXS J211336.1+542226 | 0761120801 | 39.3 | 0.65 | $21^h13^m35.38^s$ | $+54°22'32.8''$ | 1265.4 | 4 |
| CXOGC J174622.7-285218 | 0694640401 | 58.7 | 0.058 | $17^h46^m22.7^s$ | $-28°52'18''$ | 1745 | – |
| CXOGC J174517.4-290650 | 0694641001 | 47.9 | 0.008 | $17^h45^m17.41^s$ | $-29°06'50.0''$ | 321.5 | – |
| V381 Vel | 0673140401 | 14.9 | 0.10 | $10^h16^m58.90^s$ | $-41°03'44.6''$ | 7388 | 2.04 |

**Table 1.** For each source analyzed in this paper we give: the source name (Target), the *XMM*-Newton observation (ObsId), the associated exposure time in ks, the (J2000) coordinates (RA and DEC) and, when available, the WD spin (in seconds) and orbital periods (in hours) of the IPs candidates as derived from the *IPHome* database. For completeness, the observed rate (in counts s⁻¹) in the 0.2 − 10 keV energy band is given in the fourth column.

### 2.3 CXOGC J174622.7-285218

The *IPHome* target CXOGC J174622.7-285218, located at J2000 coordinates RA = $17^h46^m22.7^s$ and DEC = $-28°52'18''$, was observed by the *XMM*-Newton satellite (ObsID 0694640401) with an exposure time of ≃ 58.7 ks with the MOS 1, MOS 2 and pn cameras in full frame mode and medium filter. The observation was affected by strong flares for ∼ 7% in the three EPIC cameras, as observed in the light curves in Figure 4.

On the *IPHome* catalogue, the source is labelled as a possible IP cataclysmic variable characterized by a WD spin of ≃ 1745 s as determined by the analysis of two megaseconds *Chandra* observation Muno et al. (2009) towards the Galactic center where the source resides. The orbital periodicity is currently unknown. The *XMM*-Newton data have been already used in studies regarding the high energy view of the central region of the Milky Way (see, e.g., Ponti et al. 2015).

### 2.4 CXOGC J174517.4-290650

CXOGC J174517.4-290650 is located at J2000 coordinates RA = $17^h45^m17.41^s$ and DEC = $-29°06'50''$. It was observed by the *XMM*-Newton satellite (ObsID 0694641001) with an exposure time of ≃ 47.9 ks. All the instruments observed towards the target and MOS 1, MOS 2 and pn cameras were operated in full frame mode and with the medium filter on. The observation was not affected by flares, therefore the analysis covered the entire exposure time, as can be seen from the light curves in Figure 5.

The IP nature of the source is not clear and the *IPHome* catalogue reports a WD spin of ≃ 321.5 s (as derived by Chandra data, see Muno et al. 2003) while the orbital period is currently unknown.

### 2.5 V381 Vel

V381 Vel has J2000 coordinates RA = $10^h16^m58.90^s$ and DEC = $-41°03'44.6''$ and was observed by the *XMM*-Newton satellite (ObsID 0673140401) with an exposure time of ≃ 14.9 ks. The target was observed by the MOS 1 and MOS 2 cameras in small window mode while pn instrument was operated in large window mode. In the case of V381 Vel, the observation was affected by strong flares for most of the observation (∼ 61% of the observing time) in the three cameras. Therefore, to perform the analysis, we decided to skip the cut of bad time intervals thus avoiding to introduce too many gaps (and thus to avoid rejecting completely the observation) which would introduce spurious effects in the power spectral analysis.

The source was discovered by Greiner & Schwarz (1988) as the optical counterpart of the ROSAT All-Sky-Survey source RX J1016.9-4103 (also known as 1RXS J101659.4-410332) and recognized as an AM Her system (i.e. a CV of the polar subclass) characterized by an almost synchronously rotating binary with an orbital period of ≃ 134 minutes. By using phase resolved spectroscopy, Vennes et al. (1999) found a period of ≃ 122.5 minutes which was interpreted as the WD spin period. The same modulation of ≃ 122.5 minutes was also found by Tovmassian et al. 2004, although these authors still consider the 134 minutes as the true modulation. A detailed discussion on the nature of V381 Vel is also reported in Norton et al. (2004) where the authors interpret the low number of IPs below the period gap as an evolutionary consequence and that IPs may evolve to synchronism becoming polars. Given that the 134 minutes period seems to be a 1 day alias of the 122 modulation, it is likely that the true WD spin is ≃ 122.5 minutes. However, a clear identification is still lacking and this is the reason why we considered V381 Vel in our sample.

### 2.6 *X*-ray data reduction

Once we selected the IP candidates from the *IPHome* catalogue, we retrieved all the available *X*-ray data (see table 1 for details) from the XMM-Newton Science Archive. The observation data files (ODFs) were processed by using the *XMM*-Science analysis system (SAS version 19.0.0, Gabriel et al. 2004) and with the latest available calibration files (CCFs). With these settings, we run the SAS tasks *emchain* and *epchain* in order to produce the calibrated event files.

We further screened the data by reducing the soft proton background affecting the data (see, e.g., the analysis threads described in Ehle et al. (2008)). In this respect, we built light curves (with a bin size of 100 seconds) of MOS 1, MOS 2, and pn cameras by considering photons with energy above 10 keV where the flares (if present) are easily identified as large spikes. Hence, by requiring that the net count rate must remain below 0.35 (0.4) count s⁻¹ for the MOS (pn) cameras as recommended in Ehle et al. (2008), we identified the good time intervals (GTIs) which are then used to filter the event lists. The filtered event lists are then used in order to generate the products useful for the scientific analysis as the light curves in the soft (*S*: 0.2–2 keV), hard (*H*: 2–10 keV) and full (*S+H*: 0.2–10 keV) energy bands and the images in the full band for each *XMM*-Newton camera. In panels a-e of Figure 1, we give for each source of our sample, the resulting images as obtained by the procedure described above. Superimposed on each panel of the same figure, we show the extraction regions for the targets (solid circles) and the background (dashed circles), respectively.

For the subsequent timing analysis, and for each investigated target, we extracted the source plus background signal from a circular region centered on the nominal position of the source (see table 1). The radii[4] of the extraction regions were chosen in order to avoid nearby sources. In particular, for the source (plus background) extraction regions, we adopted radii equal to 32", 60", 40", 20" and 60", respectively for SAXJ1748.2-2808, 1RXS J211336.1+542226, CXOGC J174622.7-285218, CXOGC J174517.4-290650 and V381 Vel. When possible, the background count was estimated by using extraction regions positioned on the same chip and far from any visible source.

The light curves from the source and background regions (and for the soft, hard and full energy bands) were extracted by setting a bin size of 10 seconds (for the following quantitative analysis) and 120 seconds (only for inspecting the signal by eye). When, for a given target, the source and background light curves are extracted, these curves must be synchronized in order to start and stop exactly at the same instant of time[5]. This is a required step to be performed in order to apply the subsequent SAS script *epiclccorr* which subtracts the background from the source counts bin-by-bin and applies corrections to account for different exposure and extraction area. In particular, we produced light curves with a bin size of 10 seconds and 120 seconds (only for inspecting the signal by eye). Finally, the final MOS 1, MOS 2 and pn source (background subtracted, area and exposure corrected) light curves were averaged bin-by-bin and shifted in time in order to start from 0. As outlined in the previous Section, the timing analysis was performed on the light curves binned at 10 seconds by using the Lomb-Scargle algorithm (Scargle 1982). For each source, we tested periods between $2\Delta t$, where $\Delta t$ is the bin size of the interesting light curve, and one half of the observational window.

## 3 TIMING ANALYSIS IN X-RAYS

As described in the previous section, for each IP candidate reported in table 1 we produced the soft (S), hard (H) and full (H+S) light curves shown in Figures 2, 3, 4,5, and 6 with a bin size of 120 seconds. Note that the hardness ratio $HR$, defined here as $HR = (H-S)/(H+S)$, remains practically constant (within the associated errors) during the observation of each source.

As stressed in the introduction, one can confirm the IP nature by searching for signatures of multiple periodicities. In fact, as shown by Parker et al. (2005), such systems may be characterized by features in the periodogram at the WD spin period $P_{spin}$, the orbital period $P_{orb}$ as well as the synodic period $P_{syn}$ (given by $P_{syn}^{-1} = P_{spin}^{-1} - P_{orb}^{-1}$) and the beat period $P_{beat}$ (derived from by $P_{beat}^{-1} = P_{spin}^{-1} + P_{orb}^{-1}$) that may arise (see, e.g., Warner 1995) from the presence of different X-ray sites producing photons. However, in general, the periodogram might be characterized by features that

---

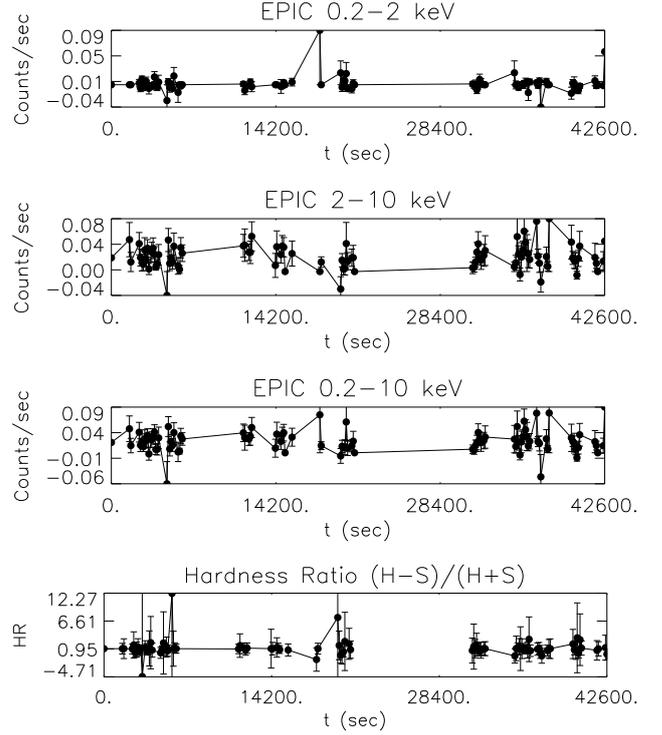

**Figure 2.** From top to bottom: the SAXJ1748.2-2808 light curves in the $0.2-2$ keV, $2-10$ keV, $0.2-10$ keV (background subtracted and synchronized) and the corresponding hardnes ratio $HR$ (see text for details).

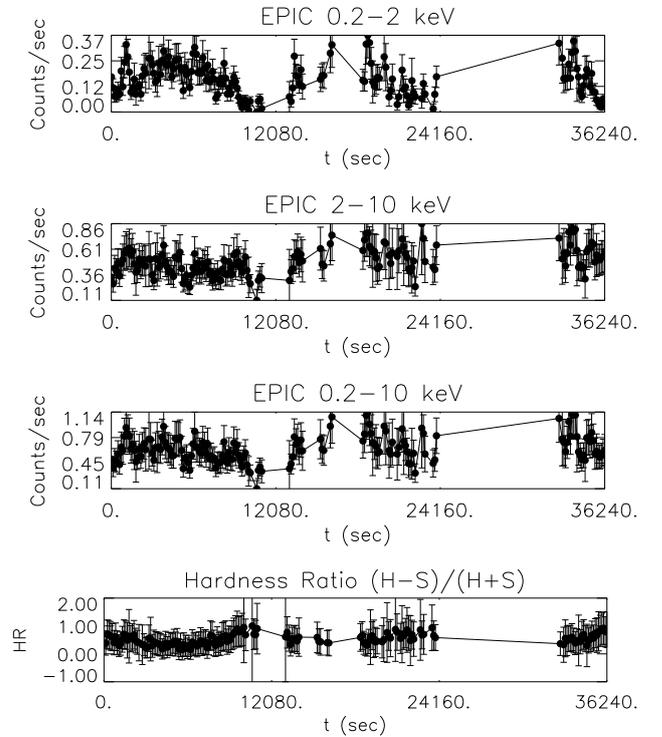

**Figure 3.** Same as in Figure 2 but for 1RXSJ211336.1+542226.

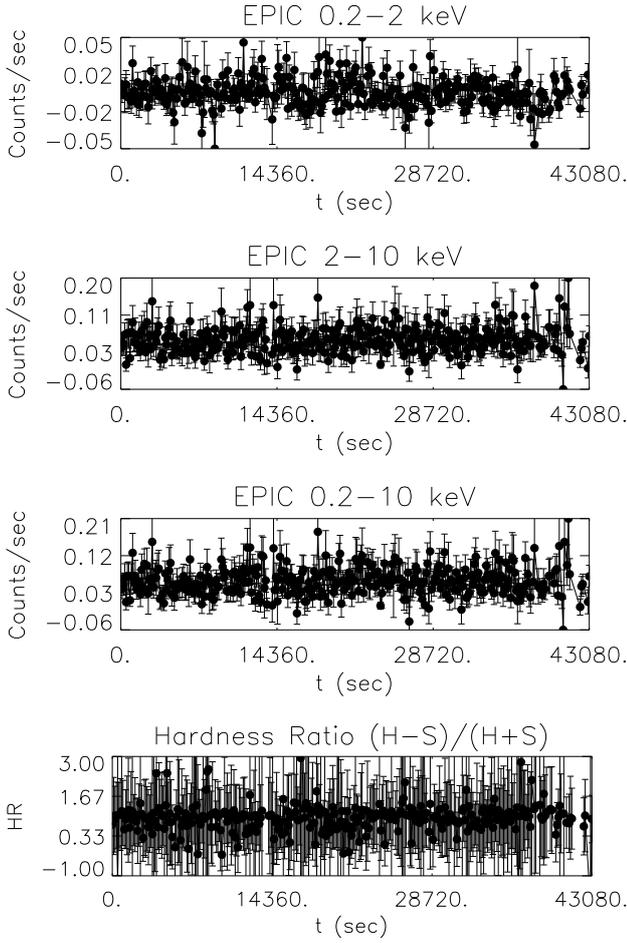

**Figure 4.** Same as in Figure 2 but for CXOGC J174622.7-285218.

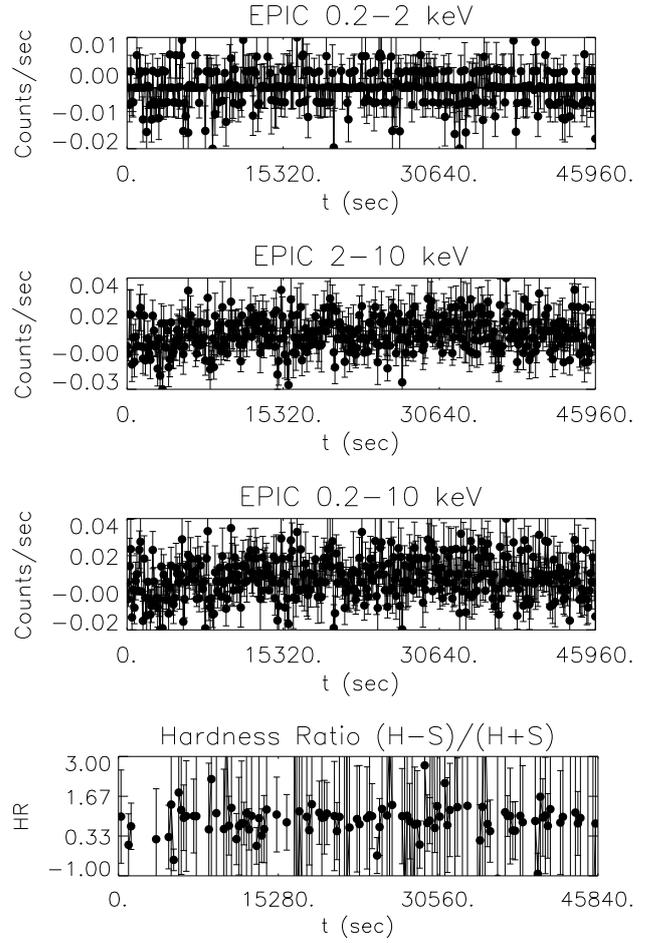

**Figure 5.** Same as in Figure 2 but for CXOGC J174517.4-290650.

could be manifest at frequencies $\omega_{i,k} = i\omega_{spin} \pm k\Omega_{orb}$, where $\omega_{spin} = 1/P_{spin}$, $\Omega_{orb} = 1/P_{orb}$, and $i$, $k$ are integers numbers. In this context, Nucita et al. (2020, 2021) and Wörpel et al. (2020) have demonstrated that recognizing the existence of peaks at pairs of values $[P_{spin}, P_{orb}]$ (together with the expected sidebands, or a few of them) is a valid tool to uncover the IP nature of these sources. In addition, one could also check for the presence of peaks at a multiple of the fundamental frequencies ($\Omega_{orb}$ and $\omega_{spin}$). In this scenario, polars (conversely to IPs) show periodograms characterized by a single prominent periodic feature and, eventually, peaks at multiples of the fundamental frequency.

Based on the previous discussion, the Lomb-Scargle periodogram associated with each IP candidate was searched for peaks corresponding to the pairs $[P_{spin}, P_{orb}]$ using, as priors, the periods (spin and/or orbital) given in the *IPHome* database. When the orbital period is unknown, a reasonable guess is done by considering $P_{orb}$ in the range 1 $P_{spin}$–1000 $P_{spin}$ (Kuulkers et al. 2006). Hence, $P_{orb}$ is considered as a good candidate value for the orbital period if *a*) a corresponding peak exists in the observed periodogram[6] and *b*) extra peaks are found at the frequencies $\omega_{i,k}$ corresponding to the expected sidebands. Of course, one could test for the presence of an infinite number of sidebands but, in practice, we tested for sidebands at the frequencies given by the integers $i$, $k$ in the range 1–12, respectively. Note that in producing the Lomb-Scargle periodogram we tested periodicities up to one half of the observational window. Therefore, in case the true orbital period is larger than this limit, the previous approach cannot be applied since any observed peak is doubtful. For these cases, the solution would be to associate any interesting peak of the periodogram to a possible sideband and to determine a posteriori the value of the orbital period with which testing other side frequencies. In this approach, an expected periodicity is associated with a feature in the periodogram only if it falls within the full width half maximum of the given peak.

### 3.1 Application to the IP candidate sample

We applied the above procedure to the periodograms of each IP candidate.

For SAXJ1748.2-2808, the expected WD spin is $P_{spin} \simeq 593$ s (i.e, 9.88 minutes) and, indeed, a large peak at $9.89 \pm 0.04$ minutes was observed. Furthermore, a relevant peak at $10.1 \pm 0.03$ minutes was observed in the Lomb-Scargle periodogram (see Figure 7, panel c). By identifying the latter as the synodic period, the expected orbital period of the system is $\simeq 545 \pm 224$ minutes, where the large uncertainty is obtained through propagation of the FWHM of the peaks centered at 9.89 minutes and 10.1 minutes. Given the huge uncertainty interval, we varied the value of the orbital period in order to maximize the number of detected sidebands. With this procedure we found that an orbital period of $P_{orb} \simeq 563$ minutes (and therefore the particular pair [$P_{spin}$, $P_{orb}$]) allowed us to predict the beat period $P_{beat} \simeq 9.72$ minutes, six harmonics of the orbital period (red dashed lines in panel a), and nine sidebands (blue dashed lines) associated with features in the Lomb-Scargle periodogram of the full light curve (see, table 2). Anyway, the peak at the proposed orbital period $P_{orb} \simeq 563$ minutes cannot be actually observed, as it is outside one half of the observational window. This detection is left to observation lasting at least $563 \times 2 = 1126$ minutes, i.e. ∼ 18 hours.

---

[6] If a peak is detected in the periodogram, we associate an error given by the Full Width Half Maximum (FWHM) of the observed feature.

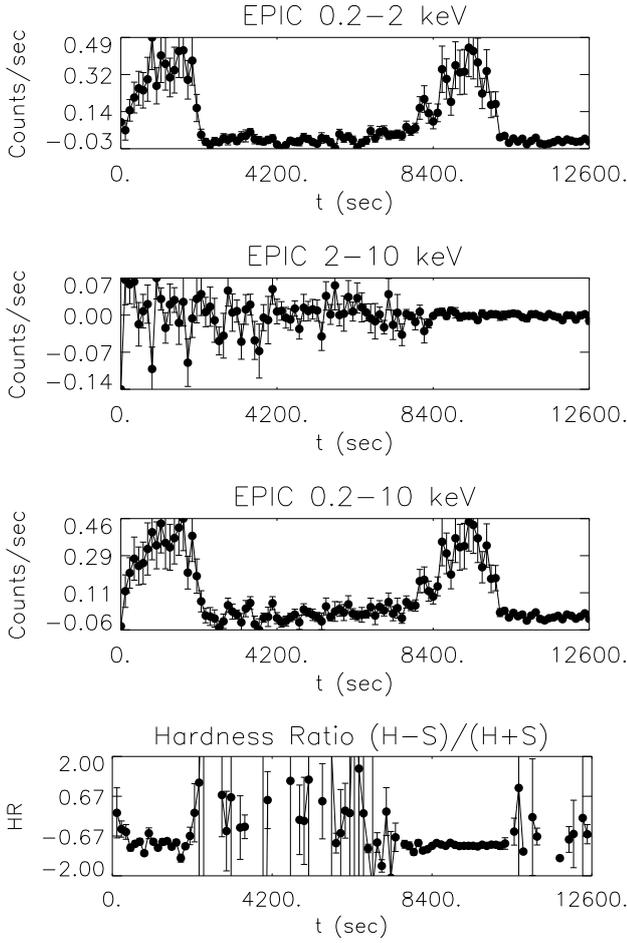

**Figure 6.** Same as in Figure 2 but for V381 Vel.

| $P_{spin}$ | $P_{orb}$ | |
|---|---|---|
| 9.89 ± 0.04 minutes | ≃ 563 minutes | |
| $P_{orb}/3 = 187.7$ | $P_{orb}/4 = 140.7$ | $P_{orb}/5 = 112.6$ |
| $P_{orb}/6 = 93.8$ | $P_{orb}/7 = 80.4$ | $P_{orb}/10 = 56.3$ |
| $P_{syn} = 10.1$ | $P_{beat} = 9.7$ | $P_{1,2} = 9.5$ |
| $P_{1,-3} = 10.4$ | $P_{2,-3} = 5.1$ | $P_{1,-4} = 10.6$ |
| $P_{1,-5} = 10.8$ | $P_{2,9} = 4.6$ | $P_{1,6} = 8.9$ |

**Table 2.** WD spin and orbital period estimated for the IP candidate SAXJ1748.2-2808. The uncertainty associated with $P_{spin}$ corresponds to the FWHM of the corresponding peak found in the periodogram of the source. We also report the theoretical values of the six detected harmonics of the orbital period, the nine expected sidebands which are effectively observed in the periodogram, and the synodic and beat periods, also found in the periodogram. All the periodicities are given in minutes. See text for details.

| $P_{spin}$ | $P_{orb}$ | |
|---|---|---|
| 21.03 ± 0.11 minutes | ≃ 250.52 minutes | |
| $P_{spin}/2 = 10.5$ | $P_{orb}/2 = 125.26$ | $P_{orb}/4 = 62.63$ |
| $P_{orb}/5 = 50.1$ | $P_{orb}/10 = 41.75$ | $P_{syn} = 22.95$ |
| $P_{beat} = 19.39$ | $P_{1,2} = 18$ | $P_{1,4} = 15.74$ |
| $P_{1,6} = 13.98$ | $P_{1,-4} = 31.65$ | $P_{2,6} = 8.4$ |

**Table 3.** Temptative WD spin and orbital period estimates for the IP candidate 1RXS J211336.1+542226. The uncertainties associated with the two periods correspond to the FWHM of the corresponding peaks found in the periodogram of the source. The values of the theoretically expected sidebands (once the reported WD spin and orbital period are assumed) which are observed and found in the Lomb-Scargle periodogram are also given (in minutes).

Note that the proposed orbital period ($P_{orb} \simeq 563$ minutes) is such that $P_{spin}/P_{orb} \simeq 0.02$, i.e. in the range of what expected for IPs. Based on this analysis we propose that SAXJ1748.2-280 is an IP candidate although the estimated orbital period is outside the observational window of the *XMM*-Newton data.

1RXS J211336.1+542226 was already analyzed by Bernardini et al. (2017) who found evidences of a WD spin and beat period of $21.09 \pm 0.24$ minutes and $22.98 \pm 0.04$ minutes, respectively, and already associated the source to the IP class. With these findings, we re-analyzed the *XMM*-Newton data of 1RXS J211336.1+542226 by applying the method described in Nucita et al. (2020, 2021), i.e. searching for the signatures of multiple side-bands (see Figure 8 panels a-e , and table 3).

We confirm the WD spin period found by Bernardini et al.

(2017) and note that the Lomb-Scargle periodogram of the 0.2-10 keV light curve shows the existence of a large asymmetric bump at ≃ 4.0 hours, where the orbital period feature is expected. Although very broad, assuming that this feature indicates the true orbital period of the system, our analysis (based on the detection of multiple sidebands and orbital harmonics) supports the hypothesis that 1RXS J211336.1+542226 is an IP.

The *IPHome* database reports, for the IP candidate CXOGC J174622.7-285218, a WD spin of $P_{spin} = 1475$ s ≃ 29.08 minutes, which is currently found by the Lomb-Scargle analysis (see Figure 9, red dashed line in panel c) as a prominent peak at $29.2 \pm 0.5$ minutes confirming the findings by Muno et al. (2009). In order to determine the orbital period of the system, we searched for other interesting peaks noting two features at ≃ 163.0 and ≃ 168.8 minutes in the soft and full light curves (see the red dashed line in panel (a) of Figure 9), respectively. Having decided to perform the analysis on the full light curve, we first tested the period pair [29.2,168.8] but no interesting sideband feature was detected. Therefore, we take the previous assumption and used the value of 163.0 minutes (observed in the soft light curve) as a prior for the orbital period. In any case, we note that within the associated error of ≃ 20 minutes as derived by the peak full width half maximum, the positions of the two features (i.e. 168.8 and 163.0 minutes) are fully consistent. Hence, we were left with six sidebands (see blue dashed lines in Figure 9, panels a-d, and table 4) which are considered detected on the basis of the full width half maximum criterion. Therefore, we assume that CXOGC J174622.7-285218 is a member of the IP subclass with $P_{spin}/P_{orb} \simeq 0.18$.

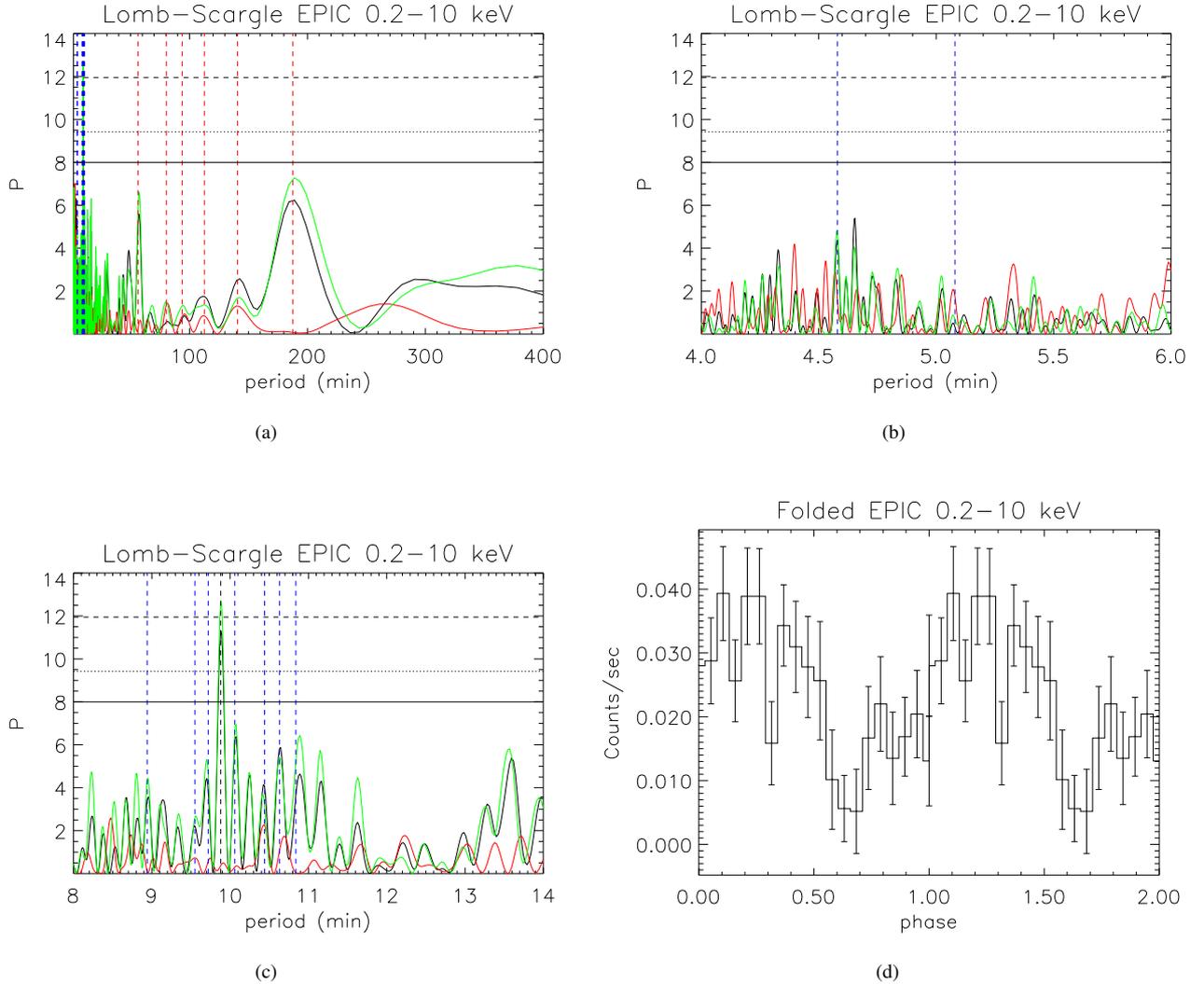

(a)

(b)

(c)

(d)

**Figure 7.** In panel (a) we give the Lomb-Scargle periodograms for the soft (red line), hard (green line) and full (black line) bands of SAXJ1748.2-280 and zooms in panels (b), and (c). Any quantitative result has been obtained using the full band. The blue dashed lines represent the positions of the side bands identified with the proposed algorithm (see text) while the red dashed ones indicate the period harmonics. The black dashed line corresponds to the WD spin period. The horizontal lines represent the 99%, 90% and 68% confidence levels. See table 2 for details on the identified harmonics and sidebands in the $0.2 - 10$ keV SAXJ1748.2-280 light curve. Finally, in panel (d) we give the light curve folded at the WD spin, clearly showing a sinusoidal pattern.

Analogously, the soft, hard and full light curves of CXOGC J174517.4-290650 have been searched for interesting periodic features. The approach outlined above allowed us to confirm the existence of a periodicity at the WD spin period $P_{spin} = 5.35 \pm 0.05$ minutes and to highlight the possible orbital period at $P_{orb} = 349 \pm 50$ minutes (see green and red dashed lines in Figure 10 and table 5). In fact, on the basis of the selected period pairs, we were able to identify the sidebands reported in table 5 and to conclude that the considered target possibly is a IP candidate with a spin-to-orbital ratio of $P_{spin}/P_{orb} \simeq 0.02$, as expected for this class of objects.

The last source in our sample is V381 Vel which was observed

| $P_{spin}$ | $P_{orb}$ |
|---|---|
| $29.2 \pm 0.5$ minutes | $163 \pm 20$ minutes |

| | | |
|---|---|---|
| $P_{spin}/5 = 5.8$ | $P_{orb}/3 = 54.3$ | $P_{syn} = 35.6$ |
| $P_{1,2} = 21.5$ | $P_{1,-3} = 63.1$ | $P_{1,3} = 19.0$ |

**Table 4.** WD spin period and orbital period proposed for the IP candidate CXOGCJ174622.7-285218, together with the theoretical values of the expected sidebands and orbital harmonics identified in the periodogram. All the periods are given in minutes. The uncertainties associated to $P_{spin}$ and $P_{orb}$ are the FWHM of their corresponding peaks observed in the periodogram.

by Greiner & Schwarz (1988) as the optical counterpart of the *X*-ray source RX J1016.9-4103 (1RXS J101659.4-410332) first identified

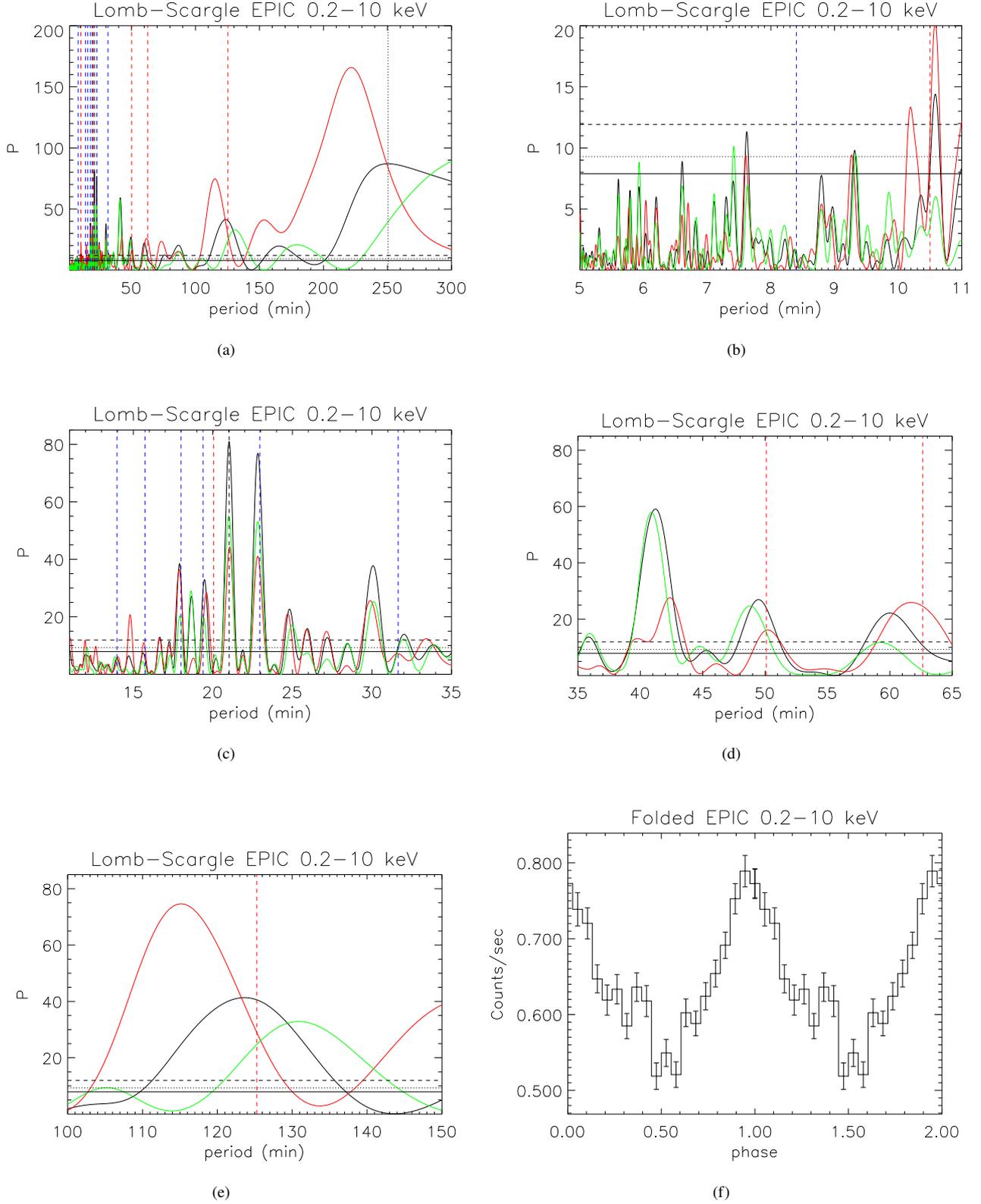

**Figure 8.** In panel (a) we give the Lomb-Scargle periodograms for 1RXS J211336.1+542226 and zooms in the panels (b), (c), (d), and (e). As in the previous figure, all the blue dashed lines represent the positions of the side bands identified with the proposed algorithm (see text). Analogously, the red dashed lines indicate the period harmonics, while the black dashed and dotted lines correspond to the WD spin and orbital period, respectively. The horizontal lines represent the 99%, 90% and 68% confidence levels. In panel (f) we give the light curve folded at the WD spin determined by our analysis (see text for details) clearly showing, as expected, a sinusoidal pattern.

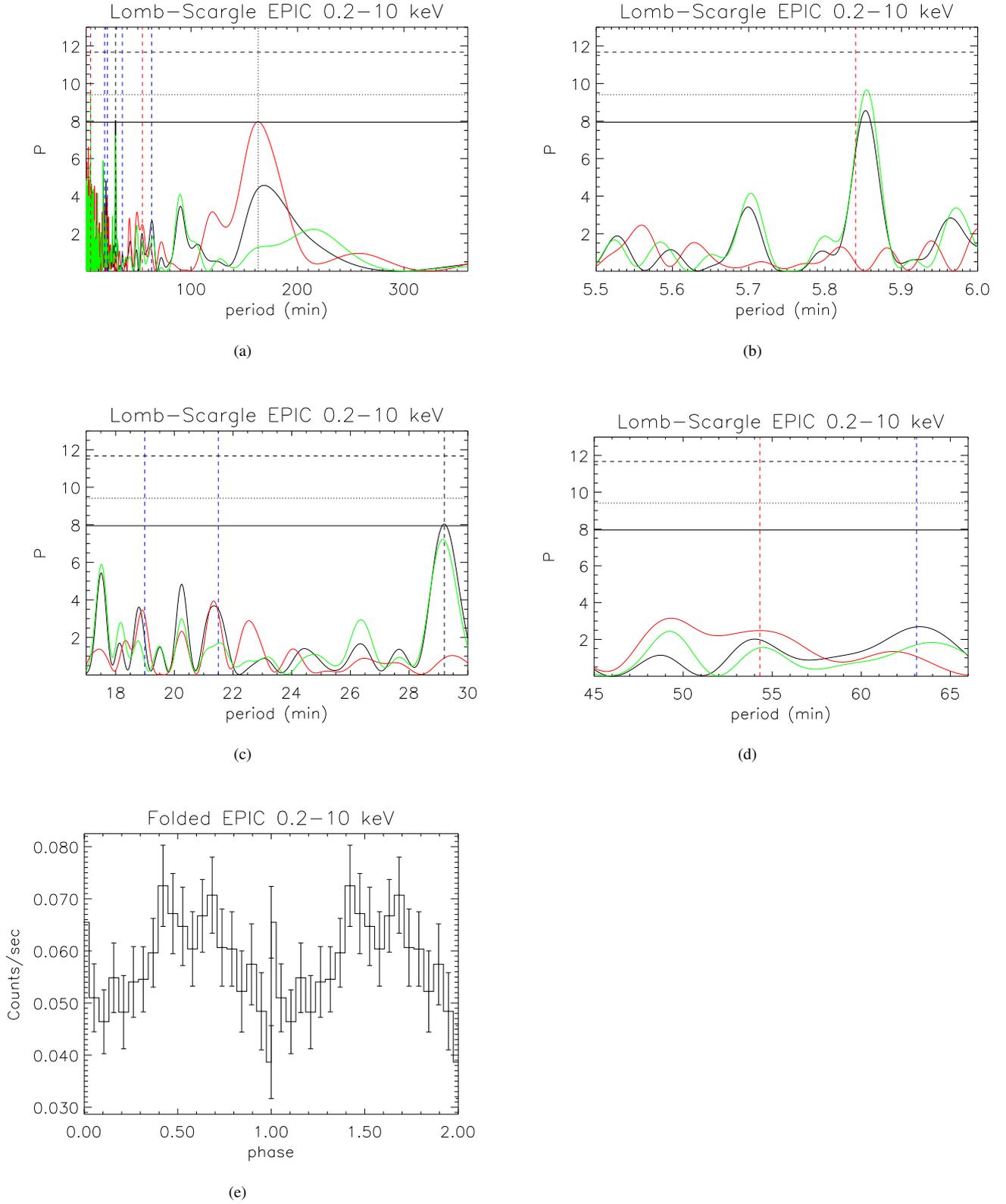

(a)

(b)

(c)

(d)

(e)

**Figure 9.** In panels (a) to (d) we give the Lomb-Scargle periodogram (and zooms) for CXOGC J174622.7-285218 flagging the possibly identified sidebands with blue dashed lines. Fundamental harmonics of the orbital period are flagged with red dashed lines. The WD spin and the proposed orbital period of ≃ 163 minutes are indicated with black vertical dashed and dotted lines, respectively. Horizontal lines give the significance at the confidence level of 99%, 90%, and 68% (see text for details). Finally, in panel (e) we give two phases of the light curve folded at the WD spin, clearly showing a sinusoidal behaviour.

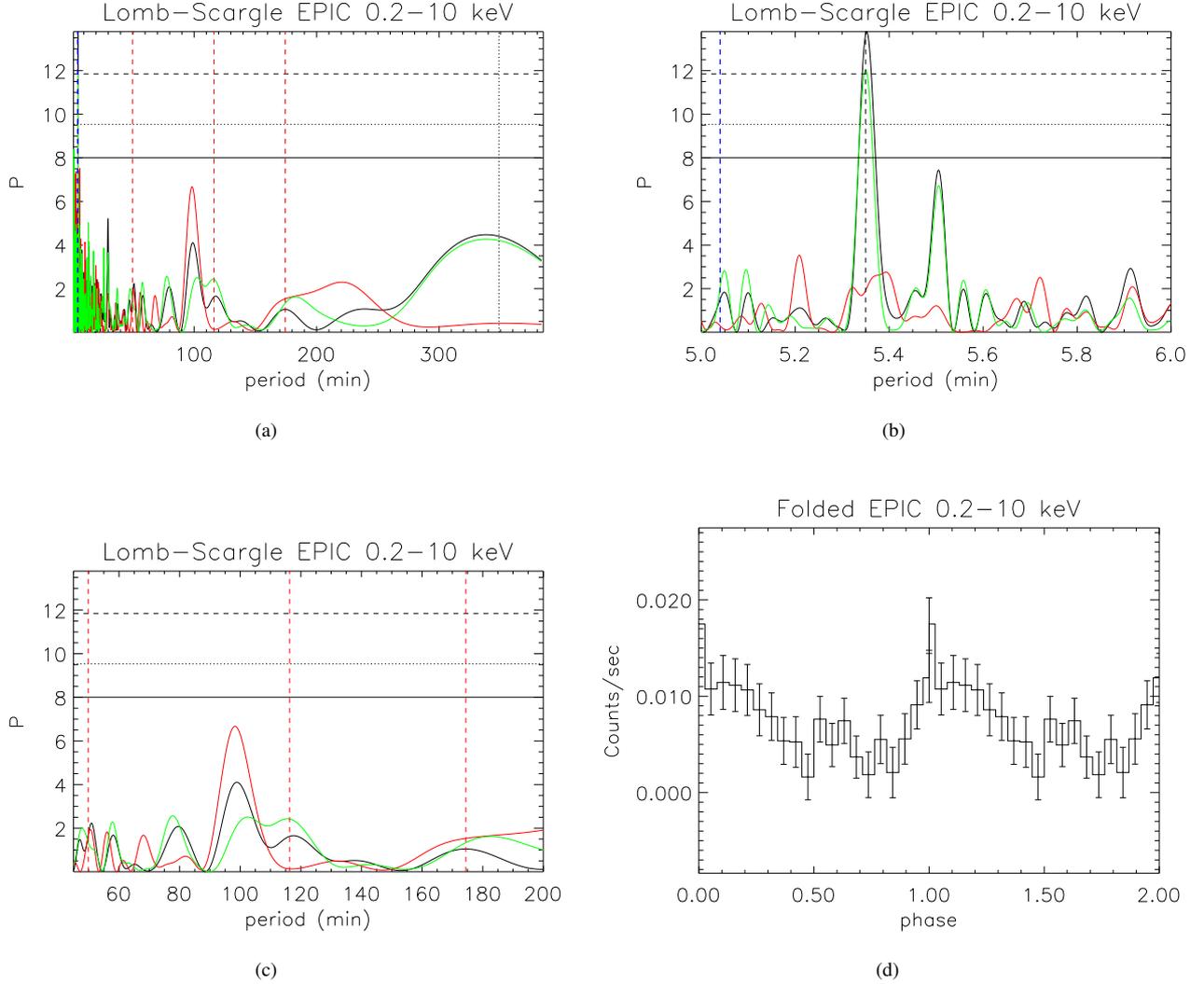

**Figure 10.** The Lomb-Scargle periodogram (panel a) and zooms for CXOGC J174517.4-290650 around interesting peaks (panel b and c). The black vertical dashed and dotted lines flag the WD spin and proposed orbital period, respectively. The identified sidebands are indicated by the blue lines while the horizontal lines give the peak significance at the confidence level of 99%, 90%, and 68% (see text for details).

| $P_{spin}$ | $P_{orb}$ | |
|---|---|---|
| $5.35 \pm 0.05$ minutes | $349 \pm 50$ minutes | |
| $P_{orb}/2 = 174.40$ | $P_{orb}/3 = 116.27$ | $P_{orb}/7 = 49.83$ |
| $P_{beat} = 5.27$ | $P_{1,4} = 5.04$ | |

**Table 5.** WD spin period and proposed orbital period (in minutes) for the IP candidate CXOGCJ174517.4-290650. Assuming these values, the timing analysis showed the existence of five sidebands, whose expected theoretical values are reported here (see text for details).

by the ROSAT All-Sky-Survey. These authors classified the target as an AM Her system (i.e. a CV of the polar subclass) with a possible orbital period of ≃ 134 minutes, slightly different than the period found by Vennes et al. (1999) of ≃ 122.5 minutes. According to the discussion in the previous section, it is likely that the true WD spin is ≃ 122.5 minutes and a clear classification of the source is still lacking. The Lomb-Scargle analysis (see, Figure 11) allowed us to identify a set of peaks (with decreasing power) at the positions (in minutes) given in table 6 which are all simple harmonics of a fundamental period $P_{spin}$ = 136 ± 16 minutes. Since we did not recognize any other interesting sideband (via the application of the described method) we conclude that V381 Vel is a polar CV with a spin/orbital period of $P_{spin}$ = 136 ± 16 minutes which given the associated error is, de facto, still consistent with the period of ≃ 134 minutes found by Greiner & Schwarz (1988).

## 4 DISCUSSION AND RESULTS

Intermediate polars are a particular class of cataclysmic variables characterized by a white dwarf accreting material from a companion star, being the binary system not synchronized. These objects are found by observing the targets in several bands as optical and X-rays and possibly searching for periodic features. In particular, the high energy signal is often characterized by modulations (Parker et al.

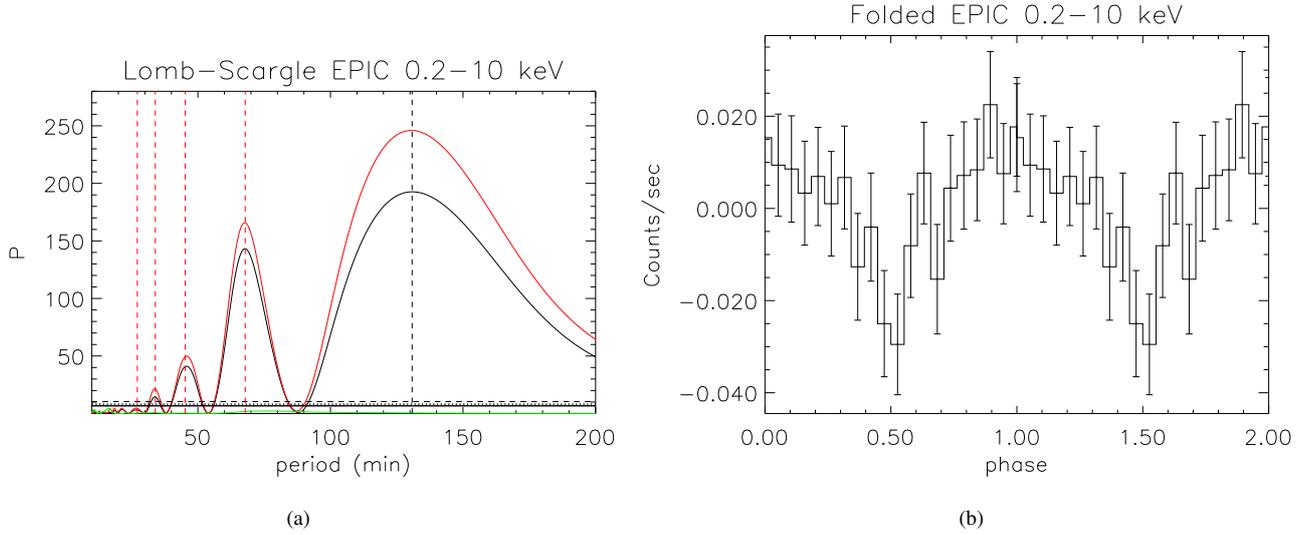

**Figure 11.** The V381 Vel Lomb-Scargle periodogram (panel a) extended up to 200 minutes and the folded light curve (panel b) at the detected periodicity of ≃ 136 minutes which shows a clear sinusoidal pattern. The black dashed vertical line represents the detected WD spin period while the dashed lines flag a few period harmonics. Horizontal lines give the significance at the confidence level of 99%, 90%, and 68% (see text for details).

| | $P_{spin}$ | $P_{orb}$ |
|---|---|---|
| | 136 ± 16 minutes | – |
| $P_{spin}/2 = 68 \pm 8$ | $P_{spin}/3 = 46 \pm 4$ | $P_{spin}/4 = 34 \pm 2$ |
| $P_{spin}/5 = 26.7 \pm 1.8$ | | |

**Table 6.** WD spin period of the polar V381 Vel and the observed simple harmonics in minutes (see text for details). All the associated uncertainties are the FWHM of the corresponding peaks in the periodogram.

2005) on the WD spin $P_{spin}$, the orbital period $P_{orb}$ as well as the synodic and beat periodicities plus a series of sidebands due to the superpositions of the signal frequencies. Therefore, recognizing the existence of multiple periodic components in the periodogram of X-ray data has been shown to be a powerful tool to classify IPs (see, e.g., Nucita et al. 2020, 2021 and Wörpel et al. 2020).

In this paper we continue the IP confirmation program of IP candidates selected from the *IPhome* database. In particular we selected the sources SAXJ1748.2-2808, 1RXS J211336.1+542226, CXOGC J174622.7-285218, CXOGC J174517.4-290650, and V381 Vel and, for each of them, we retrieved and analyzed the available *XMM*-Newton data. Then, by using the Lomb-Scargle method and the algorithm described in Nucita et al. (2020) we searched for possible signatures of the WD spin, orbital period and other sidebands in the 0.2 − 10 keV X-ray light curves.

We confirm (and/or give hints for) the IP nature for most of the source in the selected sample apart for V381 Vel which behaves like a typical CV of the polar subclass.

In the case of SAXJ1748.2-2808 (which was previously recognized as a high-mass X-ray binary (Sidoli et al. 2006) although an IP nature was not completely excluded) we found evidences that the orbital period is $P_{orb} \simeq 563$ minutes and that $P_{spin}/P_{orb} \simeq 0.02$, i.e. in the range of what expected for IPs, so that we propose to consider it as an intermediate polar binary.

The target 1RXS J211336.1+542226 was classified as an IP by

Bernardini et al. (2017). Here, based on our analysis, we support the hypothesis that 1RXS J211336.1+542226 is a member of the IP class, being characterize by a WD spin of ≃ 21.09 minutes and an orbital period of ≃ 4.0 hours.

For CXOGC J174622.7-285218, our analysis showed that the source is characterized by a spin period of $P_{spin} \simeq 29.08$ minutes (consistent with what found by Muno et al. (2009)) and give hints of the existence of an orbital period of ≃ 163 minutes. Therefore, we assume that CXOGC J174622.7-285218 is a member of the IP subclass being characterized by spin-to-orbital ratio of $P_{spin}/P_{orb} \simeq 0.18$.

As far as the source CXOGC J174517.4-290650 is concerned, our analysis allowed us to confirm the existence of a periodic feature at ≃ 5.35 minutes that, based on the sideband search, is recognized as the WD spin and we got hints of an orbital period of ≃ 349 minutes so that the source can possibly be a IP candidate with a spin-to-orbital ratio of $P_{spin}/P_{orb} \simeq 0.02$.

## ACKNOWLEDGEMENTS

This paper is based on observation from XMM-Newton, an ESA science mission with instruments and contributions directly funded by ESA Member States and NASA. We thank for partial support the INFN projects TAsP and EUCLID. The anonymous referee is acknowledged for the useful comments.

## 5  DATA AVAILABILITY

The data underlying this article were accessed from the XMM-Newton Science Archive http://nxsa.esac.esa.int/nxsa-web/#home, ID 0205240101, 0761120801, 0694640401, 0694641001 and 0673140401. The derived data generated in this research will be shared on reasonable request by the corresponding author.

This paper has been typeset from a TEX/LATEX file prepared by the author.